\documentclass[journal]{IEEEtran}

\usepackage{graphicx,subcaption}
\usepackage{amssymb}
\usepackage{makecell}
\usepackage[dvipsnames]{xcolor}
\graphicspath{{./images/}}
\usepackage{bm}
\usepackage{bbm}
\usepackage{amsmath}
\usepackage{algorithm}
\usepackage{algorithmic}
\usepackage[english]{babel}
\usepackage{graphicx}

\usepackage{background}
\usepackage{blindtext}
\backgroundsetup{scale = 1, angle = 0, opacity = 0.2,
   contents = {\includegraphics[width = \paperwidth,
   height = \paperheight, keepaspectratio]
   {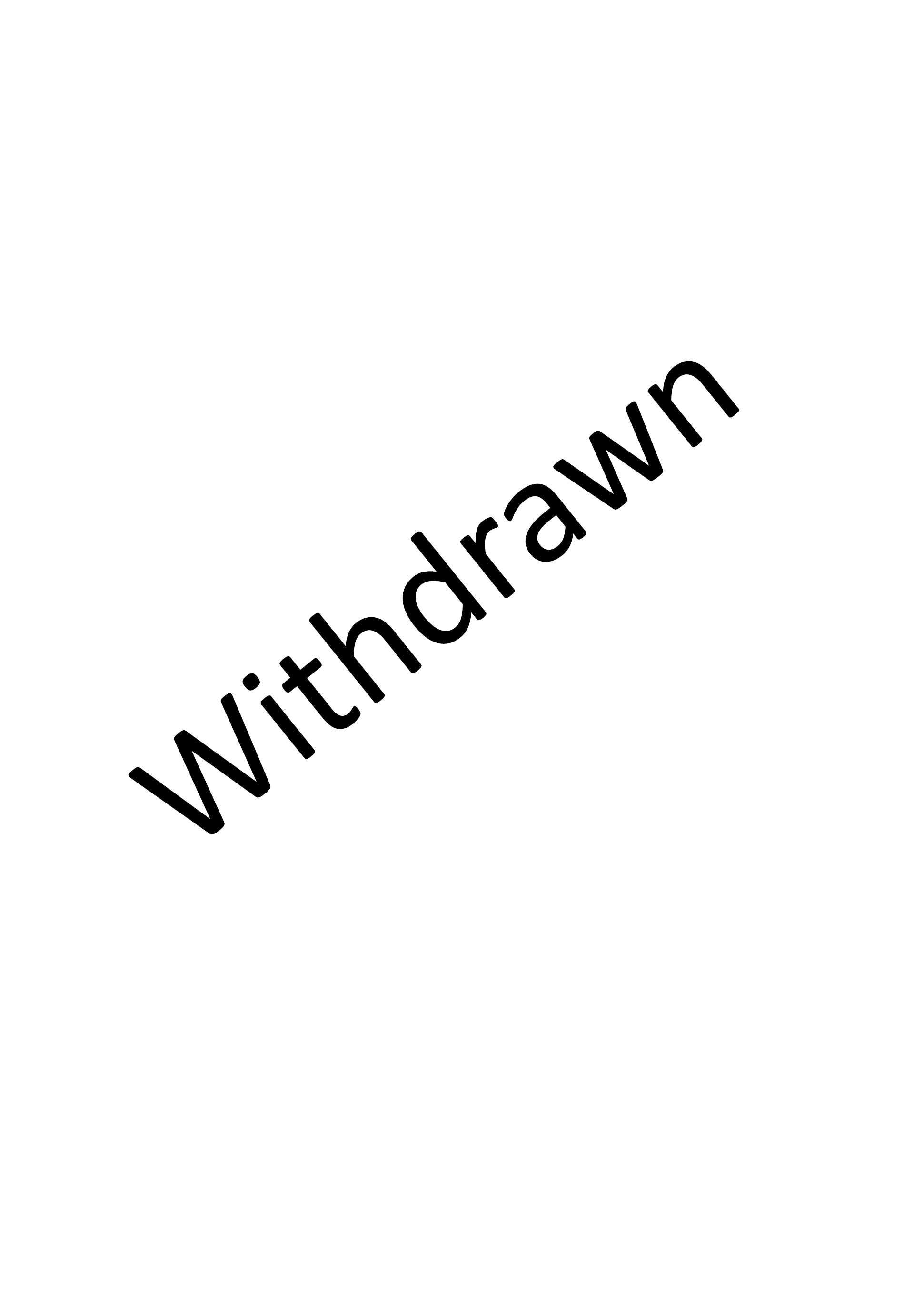}}}

\begin{document}

\title{Semi-supervised Bayesian Non-parametric Behavior Detection}

\author{Jin Watanabe, Takatomi Kubo, Fan Yang and Kazushi Ikeda}


\maketitle

\begin{abstract}
An automatic mouse behavior recognition system can considerably reduce the workload of experimenters and facilitate the analysis process. Typically, supervised approaches, unsupervised approaches and semi-supervised approaches are applied for behavior recognition purpose under a setting which has all of predefined behaviors. In the real situation, however, as mouses can show various types of behaviors, besides the predefined behaviors that we want to analyze, there are many undefined behaviors existing. Both supervised approaches and conventional semi-supervised approaches cannot identify these undefined behaviors. Though unsupervised approaches can detect these undefined behaviors, a post-hoc labelling is needed. In this paper, we propose a semi-supervised infinite Gaussian mixture model (SsIGMM), to incorporate both labeled and unlabelled information in learning process while considering undefined behaviors. 
It also generates the distribution of the predefined  and undefined behaviors by mixture Gaussians, which can be used for further analysis.
In our experiments, we confirmed the superiority of SsIGMM for segmenting and labelling mouse-behavior videos (sorry we withdraw this work).

\end{abstract}


\IEEEpeerreviewmaketitle

\section{Introduction}

Mouse behavior analysis is widely applied in biology and medical science, to investigate the effects of interventions such as gene knockout, medications and optogenetics, etc. However, in most cases, the behavior annotation highly relies on the visual inspection by experimenters, which leads to tremendous labor costs and low reproducibility~\cite{dell2014automated}, so that it has been the main bottleneck of facilitating the whole experiment process. To solve this problem, many automatic mouse behavior recognition systems have been proposed~\cite{schaefer2012surveillance, tecott2004neurobehavioral}.

In general, there are mainly two types of automatic mouse behavior recognition system, either by supervised approaches or unsupervised approaches~\cite{dell2014automated}. In supervised approaches, experimenters annotate a fraction of the whole data set by predefined behaviors. Once a classifier is trained by these labelled data, the behavior recognition can be performed automatically~\cite{jhuang2010automated,kabra2013jaaba,van2013automated}. Nonetheless, the performance of 
supervised approaches are highly dependent
on the size of labeled data, on the other hand,
there is a large supply of unlabelled data, which is not used for improving classifier performance.
Furthermore, owing to the experimental environment varies from one laboratory to another, mouses can show various types of behaviors~\cite{schaefer2012surveillance}. Therefore, besides the predefined behaviors that we want to analyze, there are many undefined behaviors existing~\cite{jhuang2010automated, burgos2012social, van2013automated}. If we cannot separate these undefined behaviors from predefined behaviors, they will affect the classier in the prediction, which means, the data belong to undefined behaviors will be assigned to predefined behaviors. Since the data belong to undefined behaviors cannot be unlabelled in advance, supervised approaches is not available to identify them.

In contrast to supervised approaches, unsupervised approaches cluster the whole data set according to the data structure, without any label information. They are especially useful when we do not know how many kind of behaviors are existing in a new experimental environment, since they have the potential to detect both predefined behaviors and undefined behaviors~\cite{berman2013mapping,braun2010identifying,brown2013dictionary}. However, without considering any label information, the performance of unsupervised approaches is hard to be improved. In addition, a post-hoc annotation for clusters is needed. 
Thus, an automated system that can incorporate both labeled and unlabelled information
in learning process while considering undefined behaviors are desirable. 

Recently, semi-supervised approaches has been receiving lots of attentions. Semi-supervised approaches not only utilizes unlabelled data, but also take the incomplete label information into consideration.
In previous work, semi-supervised approach was applied to animal behavior recognition, using accelerometer data, and its effectiveness and practicality outperform supervised learning~\cite{tanha2012multiclass}. Nevertheless, in semi-supervised learning, it is generally considered that the number of clusters is identical to the number of predefined classes. Since we want to detect undefined behaviors, the number of clusters should not be decided in advance.

As a method without specifying the number of clusters in advance, nonparametric Bayesian method has been successfully applied to analyze behavior of humans~\cite{willsky2009sharing, hughes2012nonparametric}. Besides, in case of mouse behavior analysis, Dirichlet process mixture of multinationals was applied to classifying behavior distribution~\cite{zanotto2013dirichlet}. These methods predict the number of classes while simultaneously performing model inference, however, they were applied in unsupervised approaches. 

In order to incorporate both labeled and unlabeled information
in learning process while considering undefined behaviors, we propose a semi-supervised infinite Gaussian mixture model (SsIGMM), which is an extension of infinite Gaussian mixture model (IGMM). 

In our experiments, we demonstrate the superiority of SsIGMM for identifying predefined behaviors and detecting undefined behaviors. SsIGMM is therefore can efficiently and accurately segment and label mouse-behavior videos.

\begin{figure*}[ht!]
  \begin{tabular}{rl}
   \begin{minipage}{0.5\hsize}
   \begin{center}  
   \includegraphics[height=0.65\linewidth]{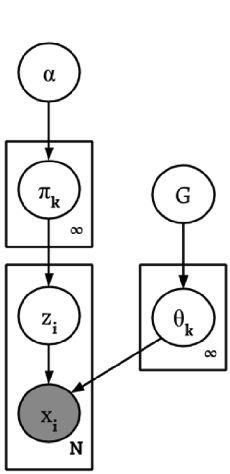}
   \hspace{0cm} (a) IGMM
  \end{center}
 \end{minipage}
 \begin{minipage}{0.5\hsize}
  \begin{center}
   \includegraphics[height=0.65\linewidth]{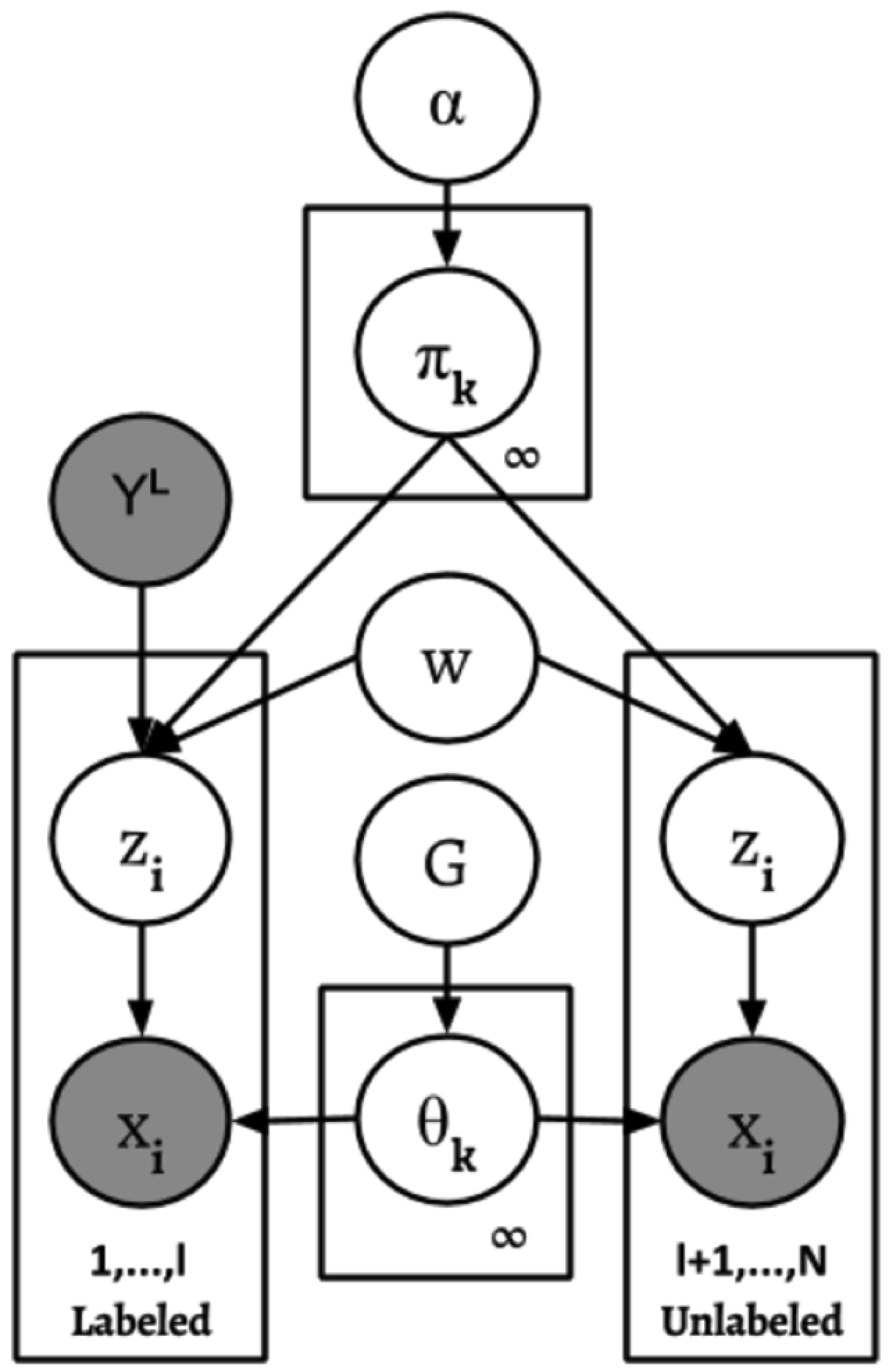}
   \hspace{0cm} (b) SsIGMM
  \end{center}
 \end{minipage}
\end{tabular}
\caption{Graphical models}
\label{fig:graphical_model}
\end{figure*}

\section{Semi-supervised extension of IGMM}

\subsection{Infinite Gaussian Mixture Model}
Since SsIGMM is developed from IGMM, we give an introduction to  IGMM before moving to SsIGMM. 

IGMM is a Nonparametric Bayesian model, its graphical model is shown in Fig.\ref{fig:graphical_model} (a). IGMM can automatically estimates the number of components with respect to the data structure, using the Dirichlet process (DP) prior~\cite{rasmussen1999infinite}. Given ${X}=\{x_1,..., x_N\}~(X \in \mathbb{R}^{N\times D})$ as the observed data set, 
there is ${Z}=\{z_1,..., z_N\}$ to indicate which Gaussian component each data is assigned to. For all Gaussian components, the set of their parameters is denoted by $\vec{\theta}=\{\theta_{1},...,\theta_{\infty}\}$, associated with their corresponding weight $\vec{\pi}=\{\pi_{1},...,\pi_{\infty}\}~(\sum_{k=1}^{\infty}\pi_{k}=1)$. For each Gaussian component, there is $\theta_{k} =\{\mu_{k},\Sigma_{k}\}$, while $\mu_{k}$ and $\Sigma_{k}$ are the mean and the covariance matrix, respectively. Thus, the generative model of IGMM is as follows:
\begin{align}
\pi_{k}  &\sim GEM(\alpha),\\
z_i &\sim  \vec{\pi}, \\
\theta_{k}  &\sim G(H) \\ 
x_i &\sim P(x_i~|~\theta_{z_i}),
\end{align}  

Where GEM is a stick-breaking distribution over $\pi$, and $\alpha$ is the concentration  parameter  \cite{teh2011dirichlet}. 
$G$ is the base distribution, where the $\theta$ is sampled from.
Here, $G$ is a Gaussian-inverse-Wishart distribution with hyper parameters $H= \{m_0,\Lambda_{0},\kappa_0,\nu_0\}$:
\begin{align}
& \Sigma \sim Inv-Wishart_{\nu_{0}}(\Lambda^{-1}_{0}) \nonumber&\\
& \mu ~|~ \Sigma \sim  N(m_{0}, \Sigma / \kappa_{0})
\end{align}
Where $m_0$ is the mean prior for $\mu$, and $\Lambda_{0}$ is the scale matrix for $\Sigma$.
The parameters $\kappa_0$ and $\nu_0$ imply how strongly we believe the prior mean and the prior covariance, respectively.
Conventionally, we have $\nu_0=D+1$ and $\kappa_0=1$. 

One popular characterization of DP is known as the Chinese restaurant process (CRP)~\cite{pitman2002combinatorial}. Conjugating Eq.(1) and Eq.(2) yields
\begin{align}
 z_i &\sim CRP(\alpha).
\end{align}

Meanwhile, owing to the conjugate property, conjugating Eq.(3) and Eq.(4) can analytically integrated out $\mu$ and $\Sigma$, getting a multivariate Student's t-distribution ~\cite{gelman2013bayesian}.
Therefore, the likelihood that $x_i$ is generated from a new component $k^{*}$ follows:
\begin{equation}
P( x_i~|~z_i=k^{*};H) \sim {\bm St}_{\nu_0-D+1}( m_{0}, \frac{\kappa_{0}+1}{\kappa_{0}(\nu_{0}-D+1)}\Lambda_{0}).
\end{equation}

Likewise, the likelihood that $x_i$ is generated from the existing $k^{th}$ component can be denoted by
\begin{equation}
P( x_i~|~z_i=k, X_{k};H) \sim {\bm St}_{\nu_k-D+1}( m_{k}, \frac{\kappa_{k}+1}{\kappa_{k}(\nu_{k}-D+1)} \Lambda_{k}),
\end{equation}
and the update equations during the inference are provided as follows:
\begin{align}
&\quad \kappa_{k}=\kappa_{0}+N_{k}\nonumber&\\
&\quad m_k=\frac{1}{\kappa_k}(\kappa_0 m_0+N_{k}\overline{X}_{k})\nonumber&\\
&\quad \nu_{k}=\nu_{0}+N_{k}\nonumber&\\
&\quad S_{k}=\sum_{i=1}^{N_k}( x_{i}-\overline{ X}_{k})( x_{i}-\overline{X}_{k})^{T}\nonumber&\\
&\quad \Lambda_{k}=\Lambda_{0}+ S_{k}+\frac{\kappa_{0}N_{k}}{\kappa_{0}+N_{k}}(\overline{ X}_{k}-m_{0})(\overline{ X}_{k}-m_{0})^{T},
\end{align}
where $X_k$ is the data in cluster $k$, $\overline{X}_{k}$ is its mean and $N_k$ is its size. $D$ is the dimensionality of $ x_{i}$. $S_{k}$ is the sum of squares matrix about the mean of component $k$. 

The posterior inference for $z_i$ can be performed by a collapsed Gibbs sampler \cite{wood2008nonparametric}. Combining Eq.(6) with Eq.(7) and Eq.(8) yields an expression that only includes indicator variables $z_{i}$. Thus, $z_{i}$ can be sampled by following equations
\begin{align}
 &  P(z_{i}=k~|~ Z^{-i},X^{-i};\alpha,H)\propto \nonumber\\
  &
  \begin{cases}
    \frac {N^{-i}_{k}}{N+\alpha-1}P( x_i~|~z_i=k, X^{-i}_{k} ;H),& \text{$k \leq K$};\\  
    \frac {\alpha}{N+\alpha-1} P( x_i~|~z_i=k^{*}; H),& \text{$k = k^{*}= K+1$} ;
  \end{cases}
\end{align}
where $X^{-i}_{k}$ is the data in cluster $k$ excluding $x_{i}$, and its size is $N^{-i}_{k}$. $K$ is the number of currently existing clusters during inference.
To calculate $P( x_i~|~z_i=k, X^{-i}_{k} ;H)$, we only need to substitute $X_{k}$ by $X^{-i}_{k}$ in Eq.(8) and Eq.(9).

\subsection{Semi-supervised Infinite Gaussian Mixture Model}

IGMM is an unsupervised learning model. However,  besides data set $X$,  it is possible to observe an incomplete label set. Here, we denote the whole label set as $Y=\{y_1,..., y_N\}$ after encoding labels to nonnegative integers. Within $Y$, the subset $Y^{L}=\{y_1,..., y_l\} ~( l<N~and~\forall y_{i}\in Y_{L}:y_{i}>0)$ represents observed labels, and subset $Y^{U}$ represents latent labels ($\forall y_{i}\in Y^{U}:y_{i}=0$).

\begin{figure}[thbp]
   \centering
   \includegraphics[width=0.98\linewidth]{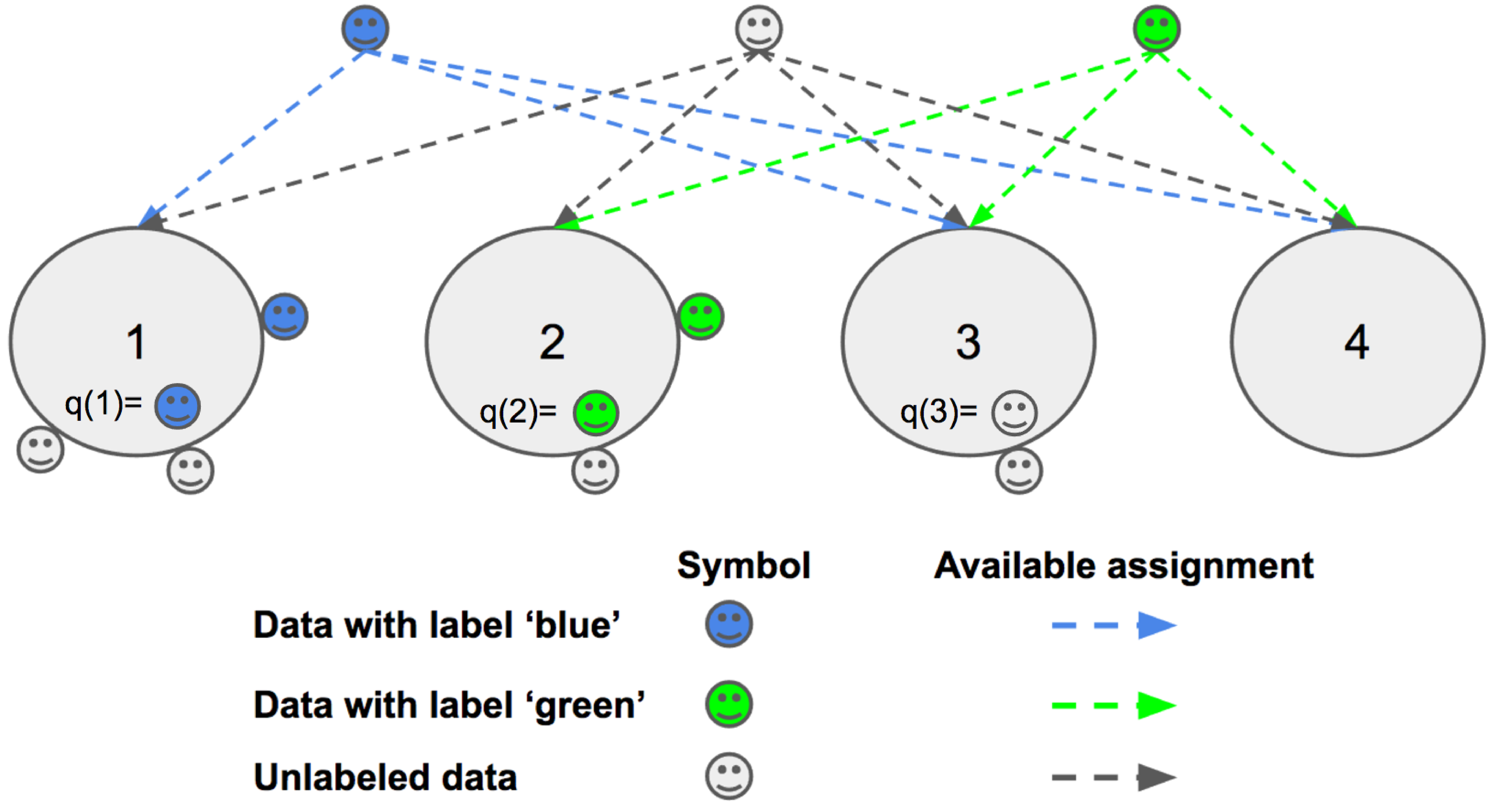}  
   \caption{\label{figure:constraint_CRP}
   Modified CRP}
\end{figure}

When estimating latent data labels, in order to improve the performance, incorporating the prior knowledge  $Y^L$ in the inference of SsIGMM is desirable. 
A generic way is to apply pairwise constraints \cite{basu2006probabilistic}, using `must-link' and `cannot-link' to indicate whether pairs of instances can belong to the same cluster or not.
Nevertheless, in order to model each distribution of predefined and undefined behavior by mixture Gaussians,
we only use the `cannot-link' constraint, so that the data with the same label can be assigned to different Gaussian components by the posterior possibility. We introduce $Q=\{q_1,...,q_{\infty}\}$ to record the observed label within each cluster, and updating $Q$ by the new assignment. Compare to IGMM, the main change is to modify CRP as Fig.\ref{figure:constraint_CRP} shows.

Hence, the inference is performed under the following rule:

\begin{itemize}
  \setlength{\itemsep}{5pt}
  \item Any pair from $Y^{L}$ cannot belong to the same component if their labels are different.
  \item If component $k$ includes unique $y_i\in Y^{L}$,  corresponding $q_{k}$ is assigned to $y_{i}$.
  \item If component $k$ does not include any $y_i\in Y^{L}$,  corresponding $q_{k}$ is assigned to $0$.
\end{itemize}

Consequently, Eq.(10) is modified as follows:
\begin{align}
 & P(z_{i}=k~|~{Z}^{-i}, X^{-i}, Y^L, Q; \alpha, H)\propto  \nonumber\\
 &
 \begin{cases}
    ~~~~$0$, 
    ~~~~\text{$k \leq K~~and~(y_{i}\in Y^L~and~y_{i} \neq q_{k}~and~q_k>0)$};\\
    \frac{N_{k}^{-i}}{N+\alpha-1}P( x_i~|~z_i=k, X^{-i}_{k} ;H), 
    ~~~~\text{$k \leq K$ and Otherwise};\\
    \frac{\alpha}{N+\alpha-1}P( x_i~|~z_i=k^*; H), 
    ~~~~\text{$k = k^{*}= K+1$} .
  \end{cases}
\end{align}

This deviation guarantees that data have different labels cannot be assigned to the same component, while data have the same label could be assigned to different components. Moreover, unlabelled data can be assigned to any component by the posterior probability. If a component only contains unlabelled data, it means we detect an undefined behavior cluster. 

We have graphical model of SsIGMM as Fig.\ref{fig:graphical_model} (b) shows. In addition, the pseudo code of Collapsed Gibbs sampler for SsIGMM is summarised as Algorithm 1.

\begin{algorithm}[h!]
\caption{~Collapsed Gibbs sampler for SsIGMM}
\label{alg1}
\begin{algorithmic}

\STATE $Y^{L}\gets \left \{ y_{1},...,y_{l} \right \} ~(\forall y_{i}\in Y^{L}: y_{i}>0)$.

\STATE Initialize components.

\FOR{\texttt{ $T$ $iterations$ }}
\FOR{\texttt{ $i =1:N$ }}

\STATE  Remove $x_{i}$'s sufficient statistics from old cluster $z_{i}$.

\FOR{\texttt{ $k =1:K$ }}
\IF{ \textit{$y_{i}\in Y^{L} $~and~$ y_{i}\neq q_{k} $~and~$q_k>0$}} 
\STATE $Posterior_{(k)}=0$.
\ELSE

\STATE  Calculate $P(z_{i}=k ~|~ Z^{-i},X^{-i}_{k};\alpha)= \frac{N^{-i}_{k}}{N+\alpha-1}$;
\STATE  Calculate  $P( x_i~|~z_i=k, X^{-i}_{k}; H) $ using Eq.(8);
\STATE  Calculate  $Posterior_{(k)}\propto $
\STATE  $P(z_{i}=k~|~Z^{-i},X^{-i}_{k};\alpha) $ $P( x_i~|~z_i=k, X^{-i}_{k}; H)$.
\ENDIF
\ENDFOR

\STATE  $k^{*} = K+1$;
\STATE  Calculate $P(z_{i}=k^{*} ~|~\alpha)= \frac{\alpha}{N+\alpha-1}$;
\STATE  Calculate  $P( x_i~|~z_i=k^*; H)$ using Eq.(7);
\STATE  Calculate  $Posterior_{(k^{*} )}\propto $
\STATE  $P(z_{i}=k^{*} ~|~\alpha)$ $P( x_i~|~z_i=k^*; H)$.    
\STATE  Sample  $k_{new}  \sim Posterior_{(1,...,k^{*} )}$ after normalizing.

\STATE If $k_{new}=k^{*}$, then $K=K+1$;
\STATE Add $x_{i}$'s sufficient statistics to its new cluster $z_{i}$=$k_{new}$.
\STATE If any cluster is empty, remove it and its related $q$, then decrease $K$.
\STATE Update $q_{k_{new}}$ to record $y_{i}$ or $0$.

\ENDFOR
\ENDFOR

\FOR{\texttt{ $k =1:K$ }}

\FOR{\texttt{ $i =1:n$ }}
\IF{ \textit{$z_{i}= k$}}
\STATE  $ z_{i}= q_{k}$ 
\COMMENT {$ q_{k} =0$ represents undefined behaviors, while $ q_{k} > 0$ represents predefined behaviors.}
\ENDIF

\ENDFOR
\ENDFOR

\end{algorithmic}
\end{algorithm}

\section{Synthetic data experiment}
\begin{figure*}[h!]
 \begin{center}
     \begin{minipage}{0.29\hsize}
       \begin{center}
         \includegraphics[clip, width=\hsize]{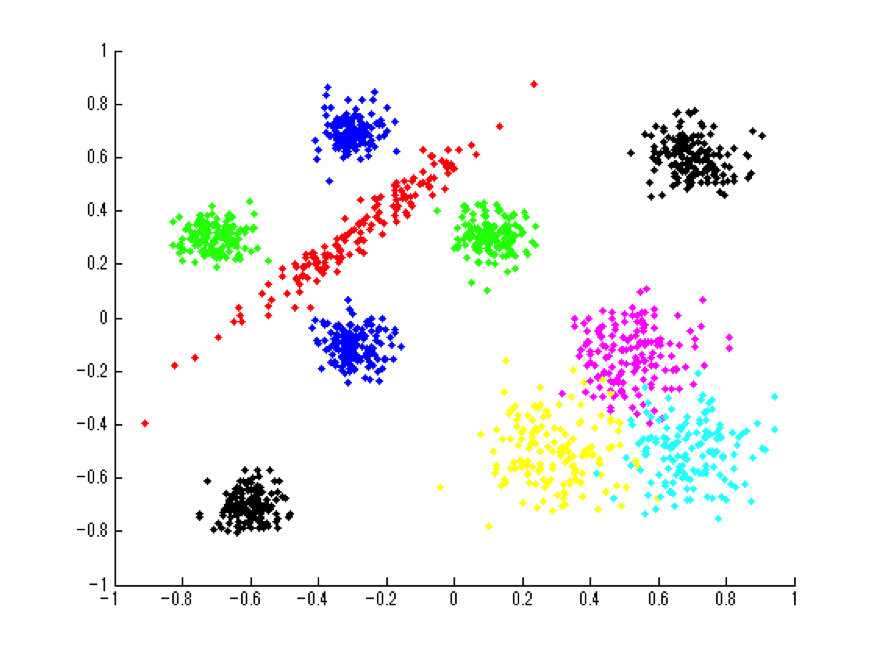}
         (a) true
       \end{center}
     \end{minipage}
     \begin{minipage}{0.70\hsize}
       \begin{tabular}{ccc}
       \begin{minipage}{0.30\hsize}
         \begin{center}
           \includegraphics[clip, width=\hsize]{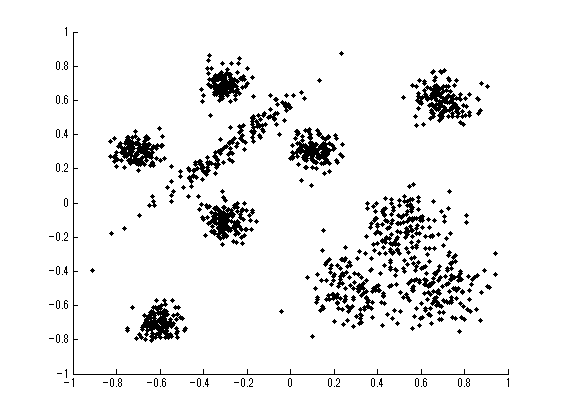}
         \hspace{-0.55cm}  (b) input (unsupervised)
         \end{center}
       \end{minipage} &
       \begin{minipage}{0.30\hsize}
         \begin{center}
           \includegraphics[clip, width=\hsize]{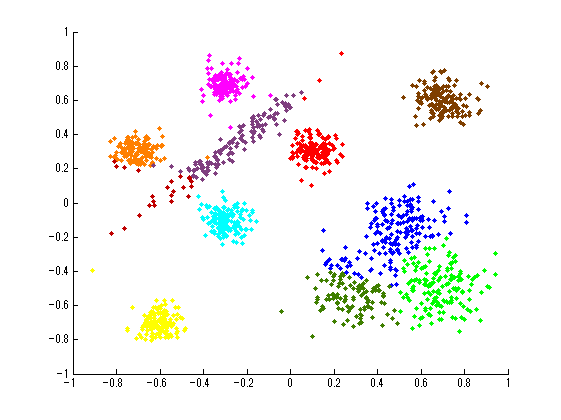}
           (c) AVDPM
         \end{center}
       \end{minipage} &
       \begin{minipage}{0.30\hsize}
         \begin{center}
           \includegraphics[clip, width=\hsize]{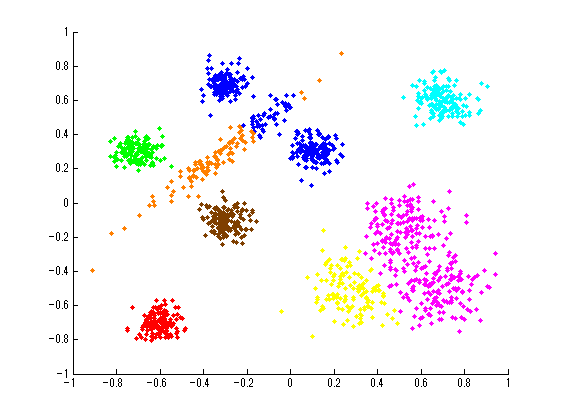}
           (d) IGMM
         \end{center}
       \end{minipage} \\\\ \hline \\
       \begin{minipage}{0.30\hsize}
         \begin{center}
           \includegraphics[clip, width=\hsize]{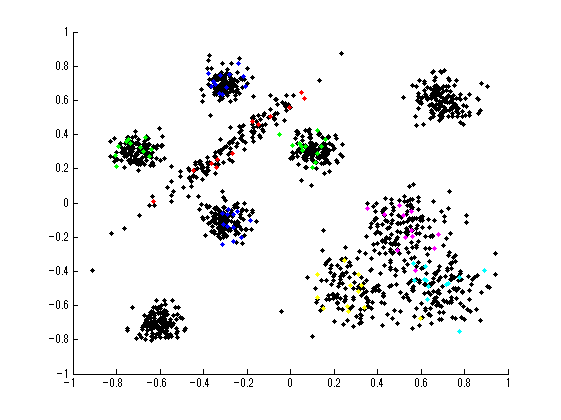}
           (e) input (semi-supervised)
         \end{center}
       \end{minipage} &
       \begin{minipage}{0.30\hsize}
         \begin{center}
           \includegraphics[clip, width=\hsize]{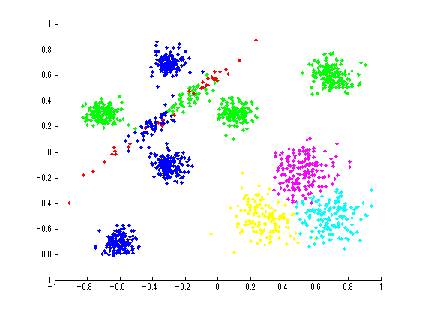}
           (f) SsGMM
         \end{center} 
       \end{minipage} &
       \begin{minipage}{0.30\hsize}
         \begin{center}
           \includegraphics[clip, width=\hsize]{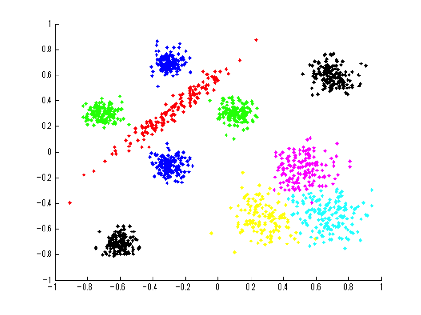}
           (g) SsIGMM
         \end{center}
       \end{minipage}
       \end{tabular}
     \end{minipage}
\end{center}
     \caption{ (a) is the true label assignment (i.e., $8$ classes). The blue and the green classes are multimodal distributions, each of them is constructed by two Gaussian components. Black dots indicate unlabelled data. (b) is the input for unsupervised learning method (i.e., IGMM, AVDPM). (e) is the input for semi-supervised learning method (i.e., SsGMM, SsIGMM), and $10\%$ of data are labeled for other Gaussian components. (c), (d), (f) and (g) are the estimated results of each method.}
     \label{figure:artificial}
\end{figure*}

We evaluate the performance of SsIGMM by artificial data first. 
In the artificial data, 1500 instances are generated by 10 two-dimensional Gaussian components (150 instances per component). We define two 
classes to be multimodal distribution, each of them is constructed by two Gaussian components. Meanwhile, for another four Gaussian components, each of them is defined as a single class. Aforementioned six classes are considered to be the predefined classes, while the last two Gaussian components are considered to be the undefined classes (Fig.\ref{figure:artificial}(a)).

We compare the performance of SsIGMM with another three methods: accelerated variational Dirichlet process mixtures (AVDPM)~\cite{kurihara2006accelerated}, IGMM and semi-supervised Gaussian mixture model (SsGMM)~\cite{zhu2007semi}. These motheds are popularly used for clustering, whilst they are similar to SsIGMM in model structure. The difference of them is shown in Table.\ref{table:model properties}. AVDPM and IGMM are both infinite mixture model trained in an unsupervised manner, while SsGMM is a finite mixture model trained in a semi-supervised manner.
    
\begin{table}[th]
  \begin{center}
  \caption{Learning manner and component number of each method}
    \vspace{0.2cm}
    \begin{tabular}{lll} \hline
      Model \hspace{0.8cm} & Learning \hspace{1.3cm} & Component number \\ \hline
      AVDPM                & Unsupervised            & Infinite           \\
      IGMM                 & Unsupervised            & Infinite         \\
      SsGMM                & Semi-supervised         & Finite         \\
      SsIGMM               & Semi-supervised         & Infinite         \\ \hline
    \end{tabular}
    \label{table:model properties}
  \end{center}
\end{table}

In this test, we use the five-fold cross validation. For unsupervised approaches (AVDPM and IGMM), all data are treated as unlabelled data. For semi-supervised approaches (SsGMM and SsIGMM), $10\%$ labelled data are randomly chosen from four training folds and holding their labels, while another $90\%$ of training dataset, coupling with the whole testing dataset, are treated as unlabelled data. This is a little different from supervised learning, which performs training and prediction, on the training dataset and testing dataset, respectively. In SsIGMM, the clustering and the prediction are performed simultaneously, such that the unlabelled testing dataset can be included in, but the evaluation only performs on testing dataset.

We applied the codes of IGMM and AVDPM which was provided by their authors. Because SsGMM code was not available, we used our own implementation by EM-algorithm. In each validation, we ran the collapsed Gibbs sampler of IGMM and SsIGMM for 2000 iterations, and the first 1500 iterations were ignored as burn-in. Because Gibbs sampler stochastically assign class label, we chose a result that has the highest posterior probability among last 500 iterations.

The clustering result is shown by Fig.\ref{figure:artificial}. 
Data which belong to undefined classes are assigned to predefined classes in SsGMM, since its number of Gaussian components is equal to the number of unique predefined classes. Furthermore, as multimodal distribution is not considered, SsGMM is therefore only use one Gaussian component to approximate all of identical labelled data, and lead to poor results. By contrast, IGMM and AVDPM can detect undefined classes, but they cannot identify the correct class structure when different classes are adjacent to each other, since label information is not utilized. 

As an evaluation metric, we used Adjusted Rand Index (ARI) ~\cite{hubert1985comparing} defined as follows:
\begin{equation}
ARI~(C,C')=\biggl\{\sum^k_{i=1}\sum^l_{j=1} \binom {m_{ij}}{2}-t_3 \biggr\}~ {\Large{\mbox{/}}}~\biggl\{1/2(t_1+t_2)-t_3\biggr\}
\end{equation}
\begin{equation}
{\rm where}~~~t_1=\sum^k_{i=1}\binom{|c_i|}{2},~~t_2=\sum^l_{j=1}\binom{|c'_j|}{2},~~t_3=2t_1t_2/n(n-1) \nonumber
\end{equation}
where $C=\{c_1, c_2,...,c_k\}$ is the true labels, while $C'=\{c'_1, c'_2,...,c'_l\}$ is the predicted labels, and $m_{ij}$ denotes the number of common data between $C$ and $C'$. ARI has its maximum score at $1$, when all of predicted labels maching true labels.

The average ARI scores for multiple evaluation runs are shown in Table.\ref{table:artificial-ARI}. Among four methods, SsIGMM achieved the highest ARI score. These results suggest the superiority of SsIGMM against another three methods.
\begin{table}[h!]
  \begin{center}
  \caption{Average of Adjusted Rand Index.}
  \vspace{0.2cm}
  \begin{tabular}{|c||c|c|c|c|} \hline
    Method & AVDPM & IGMM & SsGMM & SsIGMM \\ \hline
    ARI    & 0.72  & 0.61 & 0.62  & 0.96   \\ \hline
  \end{tabular}
  \end{center}
     \label{table:artificial-ARI}
\end{table}

\section{Mouse-behavior data experiment}
Compare to synthetic data, the real mouse-behavior data is more complicated, but our SsIGMM still can identify predefined behaviors and detecting undefined behaviors. In the following sections, we explain about the dataset acquisition, behavior-features extracting and experimental results, to demonstrate the superiority of SsIGMM for segmenting and labelling mouse-behavior videos.

\begin{figure}[h!]
  \begin{tabular}{rl}
   \begin{minipage}{0.5\hsize}
   \begin{center}  
   \includegraphics[width=0.77\linewidth]{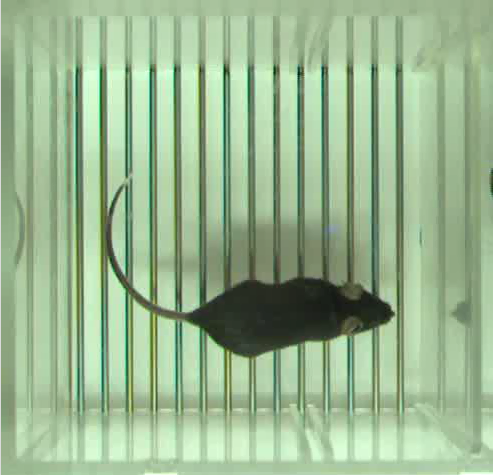}
   \hspace{1.6cm} {\footnotesize (a) Top-view}
  \end{center}
 \end{minipage}
 \begin{minipage}{0.5\hsize}
  \begin{center}
   \includegraphics[width=0.88\linewidth]{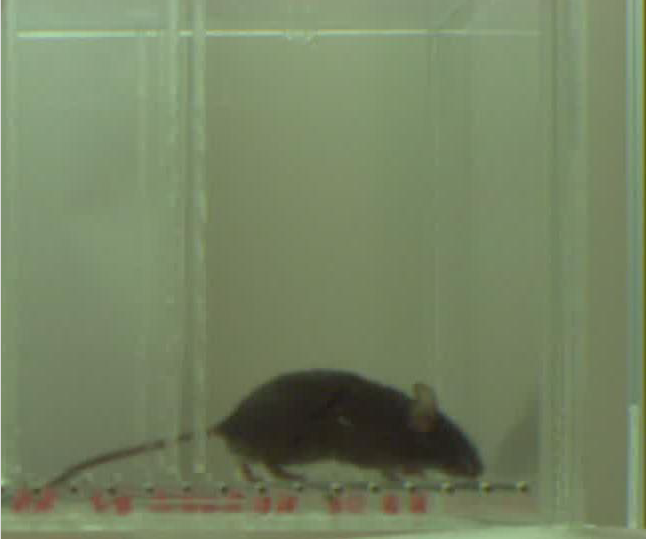}
   \hspace{1.6cm} {\footnotesize (b) Side-view}
  \end{center}
 \end{minipage}
\end{tabular}
\caption{Image examples from top and side view.}
\label{fig:view}
\end{figure}

\subsection{Dataset acquisition}
We located the experimental mouse in a transparent box, recording its activity by two synchronized cameras, from the top view and side view, respectively (see Fig.\ref{fig:view}). The recording rate is $30$ fps. All procedures were conducted under the animal welfare policies of the Institutional Animal Care and Use Committee of Nara Institute of Science and Technology. All efforts were made to minimize animals' suffering.

After recording the video, we divided mouse behaviors into six categories, including frequently observed three behaviors (walk, rear, and groom), fear related behaviors (freeze and stretch attend posture (SAP) ) and escape behavior which is induced by haloperidol or inescapable shock~\cite{anisman1981noradrenergic}.
In general, the walk, rear, and groom behaviors are considered as predefined behaviors, whereas the freeze,  SAP and escape behaviors are considered as undefined behaviors in previous studies~\cite{jhuang2010automated,van2013automated}.
We manually clipped the video by these six categories and labeled each video clips as the ground truth.

\subsection{Behavior-features extracting}

We choose following features in our model:
\begin{itemize}
  \setlength{\itemsep}{5pt}
  \item Motion of the center of gravity\hfill{}(top-view)
  \item Movement of the nose\hfill{}(top-view)
  \item Body length\hfill{}(top-view)
  \item Height of the center of gravity\hfill{}(side-view)
\end{itemize}

\begin{figure}[h!]
   \centering
   \includegraphics[width=0.5\linewidth]{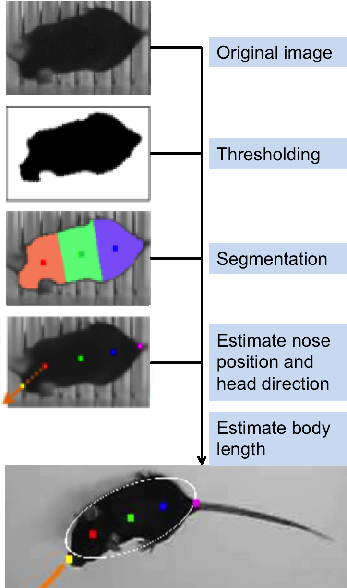}  
      \caption{\label{mouse_features_extracting_top_view}
	Extracting mouse behavior features from top-view video.}
\end{figure}
\begin{figure}[h!]
   \centering
   \includegraphics[width=0.48\linewidth]{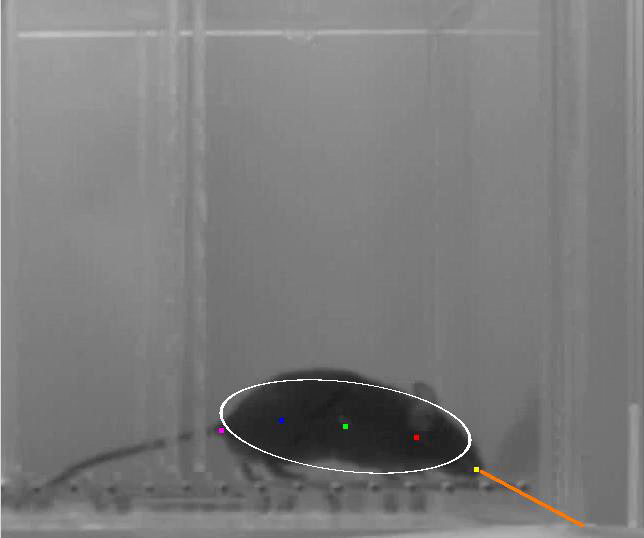}  
      \caption{\label{mouse_features_extracting_side_view}
	Extracting mouse behavior features from side-view video.}
\end{figure}

Mouse behaviors are significantly correlated to its movement. By tracking the mouse's center of gravity, we can get a robust feature which reflects its main movement. Meanwhile, the movement of nose contains important subtle information, which helps in distinguishing behaviors. Moreover, the body length and height of the center of gravity also supply critical standards to differentiate mouse behaviors. These four features are sufficient to identify different mouse behaviors in our experience. 

\begin{figure*}[ht!]
     \begin{minipage}{0.3\hsize}
       \begin{center}
         \includegraphics[clip,height=4.8cm]{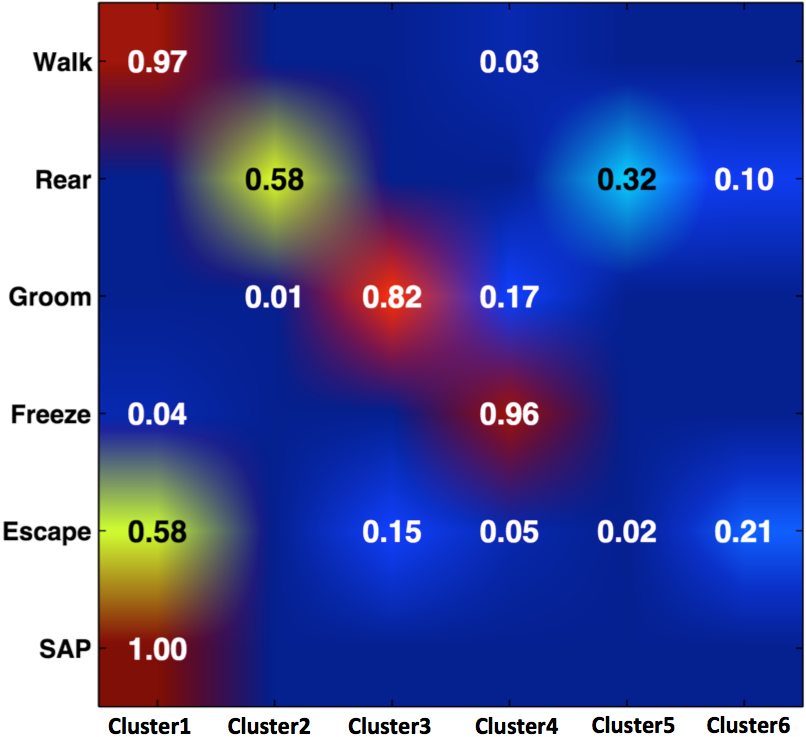}\\
         \hspace{0.2cm} (a) AVDPM
       \end{center}
     \end{minipage}
     \begin{minipage}{0.7\hsize}
       \begin{center}
         \includegraphics[clip,height=4.8cm]{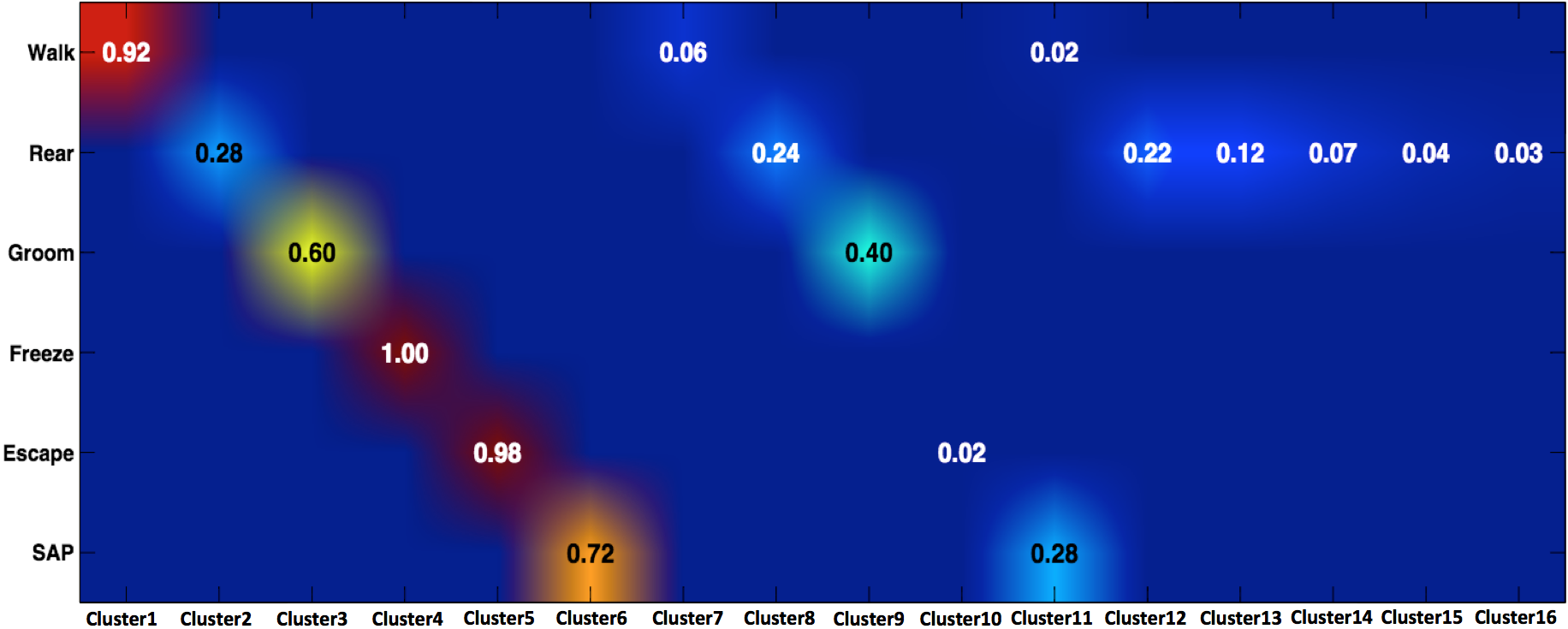}\\
         \hspace{1.5cm} (b) IGMM
       \end{center}
     \end{minipage}
\\
     \begin{minipage}{0.3\hsize}
       \begin{center}
         \includegraphics[clip,height=4.8cm]{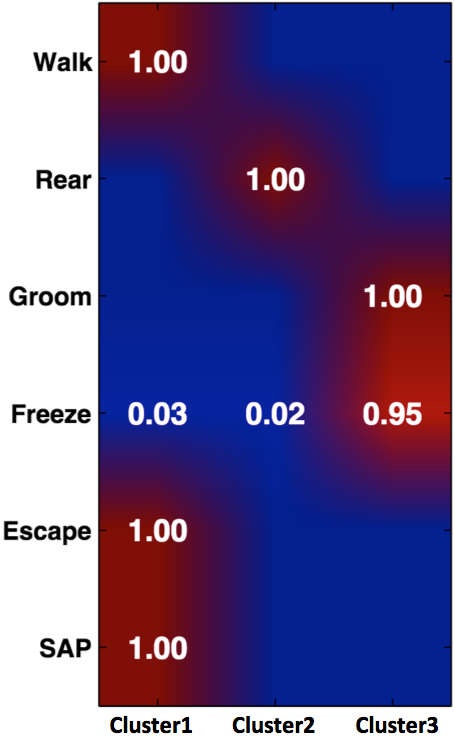}\\
         \hspace{-0.2cm}   (c) SsGMM
       \end{center}
     \end{minipage}  
     \begin{minipage}{0.5\hsize}
       \begin{center}
         \includegraphics[clip,height=4.8cm]{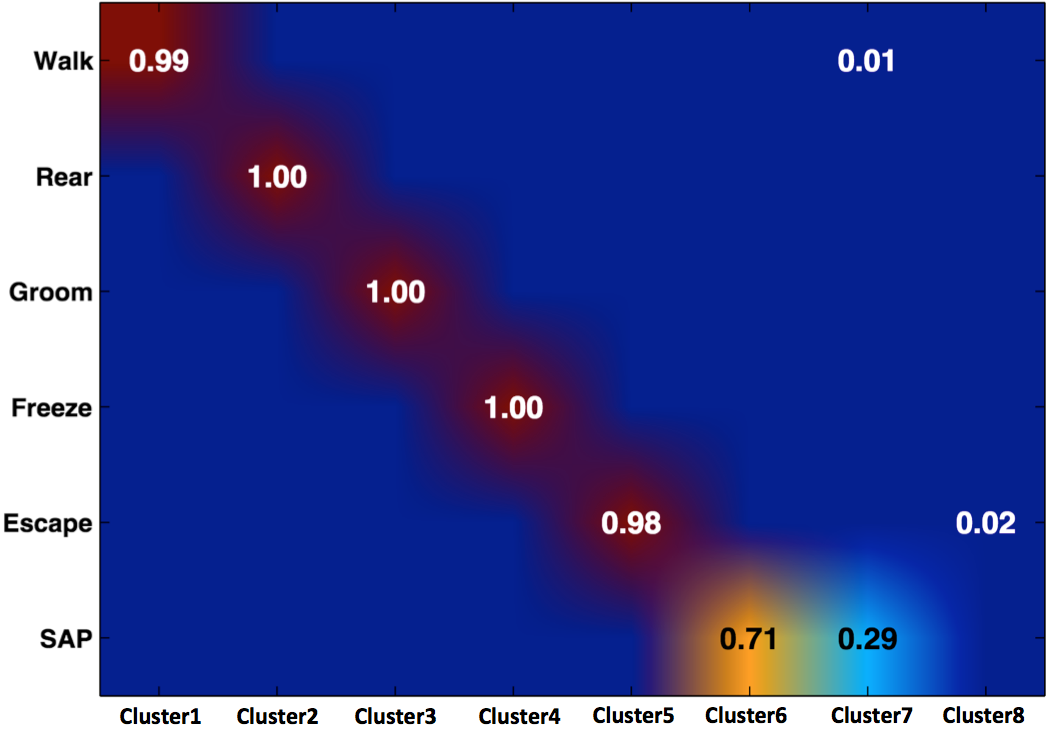}\\
         \hspace{0cm} (d) SsIGMM
       \end{center}
     \end{minipage}      
       \begin{minipage}{0.1\hsize}
       \begin{center}
         \includegraphics[clip,height=4.8cm]{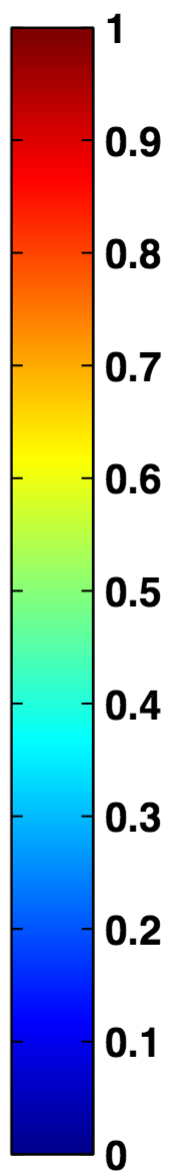}\\
         \hspace{-0.2cm}   ARI score legend \end{center}
     \end{minipage}
\\
   \caption{Confusion matrix of 1st validation. Each row indicates the behaviors, and each column indicates estimated classes by each method.}
   \label{figure:confusion}
\end{figure*}

From every top-view video frame, we can acquire the position of center of gravity, the position of nose and the body length (see Fig.\ref{mouse_features_extracting_top_view}). We first applied the thresholding method to get the silhouette of the mouse, then removed the ears and the tail. After that, we applied k-means (k = 3) to segment the mouse body into three regions: head, trunk and pelvis. We defined the horizontal position of the center of gravity is tantamount to the trunk centroid. Within the head region, the pixel which has the maximum Euclidean distance to the centroid of head is assigned to the nose.  Also, an ellipse is applied to fit the mouse body, and we assume the body length is identical to the major axis length of the ellipse~\cite{kabra2013jaaba}. Using the same method, the height of central gravity was computed from the side-view video (see Fig.\ref{mouse_features_extracting_side_view}). 
To get the movement at each frame, we calculated the average velocity in a temporal sliding window, using the corresponding position information~\cite{van2013automated}. The window size was set to $9$ frames ($\approx 1/3$ s) as previous studies applied~\cite{burgos2012social,jhuang2010automated}.

\subsection{Experimental results}

In this experiment, we also compare the performance of SsIGMM with AVDPM, IGMM and SsGMM, using five-fold cross validation. Three frequently observed behaviors (walk, rear and groom) are regarded as predefined behaviors, while another three behaviors (freeze, SAP, and escape) are regarded as undefined behaviors. For unsupervised approaches (AVDPM and IGMM), all data are treated as unlabelled data. For semi-supervised approaches (SsGMM and SsIGMM), $10\%$ labelled data are randomly chosen from four training folds and holding their labels, while another $90\%$ of training dataset, coupling with the whole testing dataset, are treated as unlabelled data (see Table.\ref{table:clipped dataset}). Other settings are the same as what we used in synthetic data experiment. 

\begin{table}[th]
   \begin{center}  
    \caption{Frame size and label information of each behavior}
    \vspace{0.2cm}
    \begin{tabular}{lrl} \hline
      Behavior \hspace{0.3cm} & Frame size \hspace{0.4cm} & Label information \\ \hline
      Walk     \hspace{0.3cm} & 716        \hspace{0.4cm} & Partially labeled ($10\%$)\\
      Rear     \hspace{0.3cm} & 956        \hspace{0.4cm} & Partially labeled ($10\%$)\\
      Groom    \hspace{0.3cm} & 903        \hspace{0.4cm} & Partially labeled ($10\%$)\\
      Freeze   \hspace{0.3cm} & 292        \hspace{0.4cm} & Unlabeled \\
      Escape   \hspace{0.3cm} & 66        \hspace{0.4cm} & Unlabeled \\
      SAP      \hspace{0.3cm} & 127         \hspace{0.4cm} & Unlabeled \\ \hline
    \end{tabular}
    \label{table:clipped dataset}
  \end{center}
\end{table}

The average ARI scores for multiple evaluation runs are shown in Table.\ref{table:mouse-ARI}.

\begin{table}[h!]
 \begin{center}
  \caption{Average of Adjusted Rand Index}
  \vspace{0.2cm}
  \begin{tabular}{|c||c|c|c|c|} \hline
    Method & AVDPM & IGMM  & SsGMM & SsIGMM \\ \hline
    ARI    & 0.68  & 0.58  & 0.79  & 0.92   \\ \hline
  \end{tabular}
  \label{table:mouse-ARI}
  \end{center}
\end{table}

We also use confusion matrix to illustrate the clustering result (see Fig.\ref{figure:confusion}). 
For AVDPM and IGMM, one behavior including several clusters does not mean a poor result, because we can perform a post-hoc notation. By contrast, if one cluster includes several behaviors by a high proportion, it represents the model clustering the data incorrectly.
In AVDPM, for instance, Cluster$1$ includes a high proportion of data from behaviors as Walk, Escape and SAP, and it is unable to distinguish them afterwards. On the other hand, IGMM does not suffer this problem, but there are too many clusters requiring a post-hoc annotation. It is inefficient in the mouse behavior recognition. Unlike unsupervised approaches, SsGMM can utilize the label information, but the number of clusters should be assigned in advance. If we follow the traditional approach to set the number of clusters equal to the categories of predefined behaviors, the result is shown in Fig.\ref{figure:confusion}.(c), The data from undefined behaviors (Freeze, Escape and SAP) are assigned to predefined behaviors. We may increase the number of clusters in this case, but the explicit number of clusters is unknown in advance.
SsIGMM solves above issues. It is able to 
cluster predefined behaviors in a high accuracy while detect undefined behaviors, and the post-hoc annotation is not needed for predefined behaviors.
Therefore, SsIGMM has the superiority for automatically segmenting and labelling mouse-behavior videos.

\section{Conclusion}
In this study, we proposed a semi-supervised infinite Gaussian mixture model (SsIGMM), which is a semi-supervised extension of IGMM. SsIGMM is able to incorporate both labeled and unlabelled information in learning process while considering undefined behaviors. It also generates the distribution of the predefined  and undefined behaviors by mixture Gaussians, which can be used for further analysis. In our experiments, we confirmed the superiority of SsIGMM for both classifying predefined behaviors and detecting undefined behaviors. Hence, SsIGMM can efficiently and accurately segment and label mouse-behavior videos.

\ifCLASSOPTIONcaptionsoff
  \newpage
\fi

\bibliographystyle{IEEEtran}
\bibliography{ref}

\end{document}